\documentclass[%
 prd,onecolumn,
 nofootinbib,
 amsmath,amssymb,
 aps,
]{revtex4-1}
\usepackage[utf8]{inputenc}
\usepackage{color}
\usepackage{amsmath,bm, physics}
\usepackage{amssymb, mathtools}
\usepackage{hyperref}
\usepackage{cleveref}
\usepackage{graphicx,xcolor}
\usepackage{dcolumn}
\usepackage{bm}
\usepackage{comment}

\newcount\colveccount
\newcommand*\colvec[1]{
        \global\colveccount#1
        \begin{pmatrix}
        \colvecnext
}
\def\colvecnext#1{
        #1
        \global\advance\colveccount-1
        \ifnum\colveccount>0
                \\
                \expandafter\colvecnext
        \else
                \end{pmatrix}
        \fi
}
\title{``Phase Coherent'' mapping of GWBs}

\date{\today}

\begin{document}


\title{Comparison of maximum-likelihood mapping methods for gravitational-wave backgrounds}

\author{Arianna I. Renzini}
\email{arenzini@caltech.edu}
\affiliation{LIGO  Laboratory,  California  Institute  of  Technology,  Pasadena,  California  91125,  USA}
\affiliation{Department of Physics, California Institute of Technology, Pasadena, California 91125, USA}
\author{Joseph D. Romano}
\affiliation{Department of Physics and Astronomy, Texas Tech University, Box 41051, Lubbock, Texas 79409-1051, USA}
\author{Carlo R. Contaldi}
\affiliation{Blackett Laboratory, Imperial College London, SW7 2AZ, United Kingdom}
\author{Neil J. Cornish}
\affiliation{eXtreme Gravity Institute, Department of Physics, Montana State University, Bozeman, Montana 59717, USA}
 
\newcommand{\air}[1]{\textcolor{orange}{[{\bf AIR}: #1]}}

\date{\today}

\begin{abstract}
Detection of a stochastic background of gravitational waves is likely to occur in the next few years. 
Beyond searches for the isotropic component of a stochastic gravitational-wave background, there have been various mapping methods proposed to target anisotropic backgrounds. Some of these methods have been applied to data taken by the Laser Interferometer Gravitational-wave Observatory (LIGO) and Virgo. Specifically, these directional searches have focused on mapping the {\em intensity} of the signal on the sky via maximum-likelihood solutions. We compare this intensity mapping approach to a previously proposed, but never employed, {\it amplitude-phase} mapping method to understand whether this latter approach may be employed in future searches. We build up our understanding of the differences between these two approaches by analyzing simple toy models of time-stream data, and run mock-data mapping tests for the two methods. We find that the amplitude-phase method is only applicable to the case of a background which is phase coherent on large scales or, at the very least, has an intrinsic coherence scale that is larger than the resolution of the detector. Otherwise, the amplitude-phase mapping method leads to an overall loss of information, with respect to both phase and amplitude. Since we do not expect these phase-coherent properties to hold for any of the gravitational-wave background signals we hope to detect in the near future, we conclude that intensity mapping is the preferred method for such backgrounds. 
\end{abstract}

\pacs{Valid PACS appear here}
\maketitle

\section{Introduction} 
Over the past two decades, several efforts have been made to understand the problem of gravitational wave (GW) map-making. 
The term ``map-making''~\cite{Cornish:2001hg,Cornish:2002bh} is typically used when considering stochastic GW backgrounds (SGWBs), which are expected to be incoherent superpositions of GW signals arriving at a detector from all directions of the sky. These GWs are generated by a variety of underlying mechanisms, which include both compact or diffuse sources at both astrophysical or cosmological distances~(see e.g., \cite{Regimbau2011, Caprini2018}).

The primary map-making method adopted in searches for anisotropy in stochastic signals is a maximum-likelihood approach that targets the overall background intensity~\cite{Allen1996}, entirely discarding any time-domain phase information present in the data. This method, hereinafter referred to as {\it intensity mapping}, has been used on cross-correlated data from the Laser Interferometer Gravitational-wave Observatory (LIGO) detectors~\citep[][]{Abadie2011,LIGO2016a,LIGOScientific:2019gaw} and, more recently, also on data from the Virgo observatory~\cite{Abbott:2021jel}. We review the basics of this method in Sec.~\ref{sec:methods}. Intensity mapping is best suited for a background that is truly stochastic. In this case the signal is characterized by waves that have random time-domain phases but whose intensity varies as a function of frequency and, in principle, angular direction. This motivates the development of methods that compress the data by discarding the time-domain phase information of the signal. For noise-dominated detectors, such as LIGO-Virgo, this is achieved by integrating the equal-time cross-correlation of the time stream of pairs of detectors. Each detector pair has a characteristic correlated sky response to the GW signal, which is time-dependent following the Earth's rotation. The set of responses determine the mapping capabilities of the detector array.

Other methods have been proposed that solve for the phase information as well, effectively mapping the amplitude and phase---or real and imaginary parts---of the two GW polarization modes on the sky~\cite{cornish2014mapping,PhysRevD.90.082001,PhysRevD.92.042003}. We refer to these as {\it amplitude-phase mapping} in what follows.
A clear motivation for preserving phase information would be the case of a diffuse, anisotropic, but {\sl coherent} GW background. It is often argued that the primordial background generated by an inflationary epoch satisfies this condition \cite{Grischuk1975,Contaldi:2018akz}. This is due to the squeezing of modes induced by any period of superhorizon evolution. However, it has been shown that, at accessible frequencies, any primordial coherence is wiped out by the effect of large-scale structure on the GWs as they propagate through the evolving universe \cite{Bartolo:2018qqn,Margalit:2020sxp}. Hence, reconstructing the phase information itself is effectively not necessary in stochastic analyses; however there remains a question of whether it is actually {\it possible} to determine the phase information for a stochastic source. This question is what we set out to answer with this paper.

In this paper we compare the application and validity of intensity and amplitude-phase mapping methods for SGWB analyses. We do this by analyzing mock data, focusing on the cross-correlation of measurements made by a network of detectors. It is important to note, at this stage, that a key element in any stochastic analysis is the accurate modeling and estimation of detector noise. We do not focus on this aspect here since our aim is to clarify the applicability of the mapping methods with respect to the signal properties. Our analysis assumes the simplest possible noise component and artificially high signal-to-noise ratios. Our conclusions, nonetheless, are independent of these choices. 

This paper is organized as follows: in Sec.~\ref{sec:GWB}, we review the strain description of the GW signal and the key assumptions made throughout the paper.
Sec.~\ref{sec:methods} details the two map-making methods.
In Sec.~\ref{sec:toys} we discuss different measurement scenarios to build an understanding of the issues encountered when reconstructing an anisotropic SGWB using intensity and amplitude-phase mapping methods. In Sec.~\ref{sec:tests} we apply the two methods to mock datasets and provide a comparison between the mapping fidelity. We conclude with Sec.~\ref{sec:discussion}, where we discuss our findings and future prospects.

\section{GW Signal}\label{sec:GWB}

The stochastic GW metric perturbations at time $t$ and position vector $\bm x$ may be written as an infinite superposition of plane waves having polarization $A$,  frequency $f$, and arriving from direction $\hat{\bm n} = (\sin\theta\cos\phi,\sin\theta\sin\phi,\cos\theta)$~\cite{Allen1996}:
\begin{equation}
    h_{ij}\,(t,\bm{x})=\int_{-\infty}^{+\infty} \!\!\!df \int_{S^2} \!\!\!d\hat{\bm n}\!\!\sum_{A=+,\,\times}\!\!h_A\,(f,\,\hat{\bm n})\,e_{ij}^A(\hat{\bm n})\,e^{-i2\pi f(\hat{\bm n}\cdot \bm{x} + t)}\,,
    \label{eq:GWstrain}
\end{equation}
where $(\theta,\,\phi)$ are the standard angular coordinates on the 2-sphere and the spatial wave vector is written explicitly as $\bm{k} = -2\pi |f| \hat{\bm n}$. We set the speed of light $c=1$ here for simplicity. The metric perturbations $h_{ij}(t,{\bm x})$ are real; hence a reality condition is imposed on the complex, frequency-domain modes, $h^\star_A(f,\hat{\bm n}) = h_A(-f,\hat{\bm n})$.
In the above expression, we choose the linear polarization basis $A=\{+,\,\times\}$, where the orthogonal polarization basis tensors $e^A$ may be written as
\begin{align}
{\bm e}^+ &=
\hat{\bm\theta}\otimes\hat{\bm\theta} - \hat{\bm\phi}\otimes \hat{\bm\phi}\,,\\
{\bm e}^\times &= \hat{\bm\theta}\otimes \hat{\bm\phi}+\hat{\bm\phi}\otimes \hat{\bm\theta}\,,
\end{align}
where
\begin{align}
\hat{\bm\theta} &= (\cos\theta\cos\phi,\cos\theta\sin\phi,-\sin\theta) \,,\\
\hat{\bm \phi} &= (-\sin\phi,\cos\phi,0)\,,
\end{align}
are the standard unit vectors tangent to the sphere.
Since $ \{h_+, h_\times\}$ are complex valued, we can write them in terms of either their real and imaginary components 
or their amplitude and temporal phase,
\begin{equation}
    h_A(f, \hat{\bm n}) = h_A^{\mathbb{R}}(f,\hat{\bm n}) + i h_A^{\mathbb{I}}(f,\hat{\bm n})= {\cal A}_A(f, \hat{\bm n}) e^{i\varphi_A(f, \hat{\bm n})}\,,
    \label{eq:phasestrain}
\end{equation}
all of which, in general, will depend on the frequency and direction of the waves.

For a stochastic background, the metric perturbations $h_{ij}(t,{\bm x})$ and hence the Fourier components $h_A(f,\hat{\bm n})$ are {\em random fields}, whose probability distributions define the statistical properties of the background.  
For the following discussion, we will assume that the background is (i) {\em Gaussian}, (ii) {\em stationary}, and (iii) {\em unpolarized}, which means that (i) the statistical properties of the random fields are completely characterized by their first- and second-order moments, (ii) there is no preferred origin of time (implying that random variables corresponding to different frequencies are statistically independent of one another), and (iii) the statistical properties of the background are invariant under rotations of the polarization tensors in the plane perpendicular to $\hat{\bm n}$ (implying statistically independent and equivalent $+$ and $\times$ polarization components). From these assumptions it follows that 
\begin{equation}
\langle h_A(f,\hat{\bm n})\rangle=0\,,
\end{equation}
that $\{h_A^{\mathbb{R}}, h_A^{\mathbb{I}}\}$ are statistically independent variants drawn from the same Gaussian distribution with zero mean and half the variance of $h_A$,
and that the phase $\varphi_A$ is uniformly distributed between $0$ and $2\pi$ and is statistically independent of the amplitude ${\cal A}_A$, which is Rayleigh distributed being the square root of the sum of squares of two statistically independent Gaussian random variables each with zero mean and equal variance.

We will further assume (iv) {\em that the background has no nontrivial phase coherence across the sky.} 
This means that the phase of the GWB signal coming from two different directions on the sky are statistically independent of one another\footnote{We define the covariance of two complex variables $A$ and $B$ as ${\rm Cov}(A,B)\equiv \langle AB^*\rangle-\langle{A}\rangle\langle{B^*}\rangle$.}:
\begin{equation}
{\rm Cov}\left[e^{i\varphi_A(f,\hat{\bm n})}, e^{i\varphi_{A'}(f',\hat{\bm n}')}\right]\propto\delta_{AA'}\delta(f-f') \delta(\hat{\bm n},\hat{\bm n}')\,.
\end{equation}
Although primordial backgrounds, such as those generated during an epoch of inflation, may be phase coherent as modes reenter the horizon and begin to propagate, this coherence is lost through propagation effects~\cite{Margalit:2020sxp}. It is therefore 
reasonable to assume that phase incoherence is a generic feature of {\it any} stochastic background that may, or may not, have angular correlations in amplitude.
In other words, the intrinsic angular scale $\Delta\hat{\bm n}$ over which the phases are correlated goes to 0, even though the correlation scale for the amplitude may be finite.
Hence, any attempt to measure the GW phase using a detector with finite angular resolution will necessarily {\em average} the true phase over this angular resolution scale.  This loss of information will degrade reconstruction of the amplitude if one tries to estimate it from the real and imaginary parts of the Fourier components.
We will show this explicitly in Secs.~\ref{sec:toys} and \ref{sec:tests}.

Putting together all of the above results, we can write
\begin{equation}
\langle h_A^{}(f,\,\hat{\bm n}) h^\star_{A'}(f',\,\hat{\bm n}')\rangle = \frac{1}{2}I(f,\,\hat{\bm n})
\,\delta_{AA'}\delta(f-f')
\delta(\hat{\bm n},\hat{\bm n}')\,\,,
\label{eq:stokey}
\end{equation}
where 
\begin{equation}
I(f,\hat{\bm n})\equiv
\frac{2}{T}\sum_A \langle{\cal A}^2_A(f,\hat{\bm n})\rangle
\end{equation}
defines the intensity of the GWB as a function of frequency $f$ and direction $\hat{\bm n}$. Here $T$ is the total observation time.
The angle brackets  $\langle\dots\rangle$ denote {\em ensemble} averaging over the random amplitudes and phases of the Fourier coefficients of the metric perturbations at a fixed spatial location $\bm x$, under the assumption of time stationarity.
In practice, this averaging is realized by averaging over all the available GW time-series data, assuming that the background is {\em ergodic}.
The distribution of energy and matter in the universe, e.g., large-scale structure, which gives rise to the GW background is {\em fixed} with respect to this averaging process.

The GW background may be {\em isotropic}, {\em anisotropic}, or {\em statistically isotropic} (i.e., invariant under arbitrary rotations of the sky) depending on the statistical properties of $I(f,\hat{\bm n})$ with respect to the sky direction $\hat{\bm n}$.  
For example, for both isotropic and anisotropic backgrounds, the intensity field is a {\em deterministic} quantity; it is independent of sky direction for a purely isotropic background---i.e., $I(f,\hat{\bm n})\equiv I(f)$, and has preferred directions for an anisotropic background.  For a statistically isotropic background, the intensity is a random field, assumed here to be approximately Gaussian, whose mean is independent of sky direction
\begin{equation}
\langle I(f,\hat{\bm n})\rangle_\Omega \equiv I(f)\,,
\end{equation}
and whose quadratic expectation values depend only on the angular separation between two points on the sky as
\begin{equation}
    \expval{I(f,\,\hat{\bm n})I^\star(f,\,\hat{\bm n}')}_\Omega \equiv C (f,\hat{\bm n} \cdot \hat{\bm n}') = \sum_{\ell = 0}^\infty \frac{2\ell + 1}{4\pi} C_\ell(f) P_\ell(\hat{\bm n} \cdot \hat{\bm n}')\,,
\end{equation}
where $P_\ell$ is the Legendre polynomial of order $\ell$. 
In the above expressions, $\langle\dots\rangle_\Omega$ denotes averaging over different GW universes~\cite{Jenkins:2019nks} (e.g., over different realizations of large-scale structure) which are drawn from a rotationally invariant probability distribution described by the {\em angular power spectrum} $C_\ell(f)$.
We note that there is a subtle difference if one imposes statistical isotropy on the $h_A$ fields; for more details regarding this see~\cite{PhysRevD.90.082001}.

In all cases, the intensity $I(f,\hat{\bm n})$ may be related to the fractional energy density parameter $\Omega_{\rm GW}(f,\hat{\bm n})$ via~\cite{Allen1996}
 \begin{equation}
	\Omega_{\rm GW}(f,\hat{\bm n}) = \frac{4\pi^2f^3}{\rho_c G}I(f,\hat{\bm n})\,,
	\label{eq:omegatoI}
\end{equation}
which is the fundamental relation that allows one to connect GWB observations to the cosmological implications of the background.
Integrating the above equation over direction on the sky yields \begin{equation}
	\Omega_{\rm GW}(f) = \frac{4\pi^2f^3}{\rho_c G}I(f)\,,
\end{equation}
which relates the monopole components of the fractional energy density parameter and the intensity of the background.
Finally, we note that it is common in the literature to assume that the intensity factorizes as
\begin{equation}
    I(f,\hat{\bm n}) = E(f) I(\hat {\bm n})\,,
\label{eq:freq_dependence_factorised}
\end{equation}
where $E(f)$ and $I(\hat{\bm n})$ encode the spectral and directional dependence of the background, respectively. We employ this assumption throughout as it considerably simplifies map-making.

\section{Map-Making methods with GW detector data}\label{sec:methods}

In this section we present two proposed maximum-likelihood map-making methods for GWBs in a common formalism. First we review the intensity mapping method which has been previously presented in several references, e.g.,~\cite{Allen1996,Romano2017,Renzini:2018vkx}, and applied consistently to LIGO data~\cite{Renzini2019a,Renzini2019b,LIGO2016,LIGO2016a,LIGOScientific:2019gaw,Abbott:2021jel}; then, we lay out the amplitude-phase mapping method which follows the presentation in~\cite{PhysRevD.92.042003}.

\subsection{Intensity mapping} 
Intensity mapping, such as that used for map-making with the LIGO-Virgo detectors, works with the cross-correlation of time-coincident data directly to discard the autocorrelated noise terms which would otherwise dominate the calculation. 

To start, we consider a set of detectors $i = \{1, \dots, N \}$. The data collected by detector $i$ in a time segment $\tau$ may be considered as made up of separate signal and noise components,  $d^\tau_i(t) = s^\tau_i(t)+n^\tau_i(t)$. We write the corresponding discrete Fourier transform as 
\begin{equation}
    \tilde{d}^\tau_i(f) = \tilde{s}^\tau_i(f)+\tilde{n}^\tau_i(f) \rightarrow \bm d^{\tau,f}\,,
    \label{eq:data_f}
\end{equation}
where we have dropped the tilde and detector label in favor of the more concise boldface vector notation. As we are dealing with a real time-stream Fourier transformed into a discretized Fourier space, it is convenient to relabel frequencies as a discrete index, which picks out a single frequency mode in the transform.
The signal component in the data is modeled as
\begin{equation}
     \bm s^{\tau,f} = \int_{S^2} d \hat{\bm n} \sum_{A}\bm R_{A}^{\tau,f}(\hat{\bm n})\,h_{A}^f(\hat{\bm n})\,,
    \label{eq:strain_model}
\end{equation}
where $\bm R$ is the {\it response} function spanning detector space. Note that the $h$ field has no $\tau$ dependence as it is assumed to be stationary.
A pair of detectors $i, j$ then observe $d_i^{\tau,f}$, $ d_j^{\tau,f}$ respectively; these form a \emph{baseline} and we can consider a correlated data vector as spanning the space of different baselines directly, $D^{\tau,f}_{ij} \equiv d_i^{\tau,f} d_j^{\tau,f\star}\equiv \bm D^{\tau,f}$. We can write down the likelihood for the residuals of the cross-correlated data as
\begin{equation}
    \mathcal{L} \propto \prod_{\tau, f, ij} \frac{1}{|\bm C_N|^{1/2}} e^{-\frac{1}{2}\qty(\bm D-\bm S) \, \bm C_N^{-1} \, \qty(\bm D -\bm S)^\star } \,.
\label{eq:LIke}
\end{equation}
Note that we are using shorthand notation here, omitting $\tau$ and $f$ everywhere in Eq.~(\ref{eq:LIke}). As in~\cite{Renzini:2018vkx},
we take the signal model $\bm S$ for the cross correlation to be
\begin{equation}
    \bm S^{\tau,f} = \int_{S^2} d \hat{\bm n}\, \bm \Gamma^{\tau,f}(\hat{\bm n}) I(\hat{\bm n})\,,
    \label{eq:crossdata_model_pix}
\end{equation}
which may be derived directly from Eq.~\eqref{eq:stokey}. Here $\bm \Gamma$ is the cross-correlated response vector to the GW intensity, obtained via the squared sum of the response terms above as
\begin{equation}
    \Gamma^{\tau, f}_{ij} (\hat{\bm n}) \equiv  \sum_A R^{\tau, f}_{A i} (\hat{\bm n})\, R^{\tau, f \star}_{A j} (\hat{\bm n})\,.
\end{equation}
The spectral dependence of $I$ has been assumed factorizable here and absorbed into $\bm\Gamma$ for simplicity, assuming Eq.~(\ref{eq:freq_dependence_factorised}). 
The time dependence (encapsulated in $\tau$) and spatial dependence of the response function are fundamental to map-making, as they define the scan strategy of the set of baselines, which sets the resolution of the measurement. More details on this may be found in~\cite{Cornish:2001hg,Taruya2005,Ballmer2006,Taruya2006,Mitra2008}.

The noise covariance matrix may be written in baseline space, $N^{\tau,f}_{ij} \equiv n^{\tau,f}_i n^{\tau,f\star}_{j} \equiv \bm N^{\tau,f}$, hence the noise covariance matrix is 
\begin{equation}
    \bm C^{\tau,f}_N = \langle \bm N^{\tau,f} \otimes \bm N^{\tau,f\star} \rangle \equiv \text{diag}(P^{\tau,f}_i P^{\tau,f}_j)
\end{equation}
assuming uncorrelated noise between detectors, where $P_i$ is the (two-sided) noise power spectrum in detector $i$. 
Maximizing the likelihood~(\ref{eq:LIke}) above yields the mapping equation for GW intensity
\begin{equation}
    I(\hat{\bm n}) = M(\hat{\bm n},\hat{\bm n}')^{-1}\,z(\hat{\bm n}')\,.
\end{equation}
We refer to $z$ as the {\it projection} of the dataset into pixel space, while $M$ is the {\it Fisher matrix} of the mapping problem. These are constructed from the following quantities calculated at the individual $f$ and $\tau$, as each of these constitutes an independent measurement according to the likelihood in Eq.~\eqref{eq:LIke},
\begin{equation}
        \bm z^{\tau, f}(\hat{\bm n}) = \bm \Gamma^{\tau, f}(\hat{\bm n})\,\qty(\bm C^{\tau, f}_N)^{-1}\, \bm D^{\tau, f}\,,\qquad
        \bm M^{\tau, f}(\hat{\bm n},\hat{\bm n}') = \bm \Gamma^{\tau, f}(\hat{\bm n})\,\qty(\bm C^{\tau, f}_N)^{-1}\,\bm \Gamma^{\tau, f}(\hat{\bm n}')\,,
    \label{eq:Imapsol}
\end{equation}
which are then summed over all times, frequencies, and baselines $ij$ in the set to obtain the full (and most informative) projection and Fisher matrix,
\begin{equation}
        z(\hat{\bm n}) = \sum_{\tau, f, ij} \bm z^{\tau, f}(\hat{\bm n})\,,\qquad
        M(\hat{\bm n},\hat{\bm n}') = \sum_{\tau, f, ij} \bm M^{\tau, f}(\hat{\bm n},\hat{\bm n}')\,.
\end{equation}
Note that the integration over frequencies here requires an assumption for the frequency dependence of the GWB, which we have transferred to the response function. One could also choose to {\it not} integrate over frequencies, and solve for maps mode by mode, avoiding this frequency modeling step.

At this stage, it is necessary to pick a working resolution on the sky in order to explicitly carry out the calculations. Using the {\tt healpix Python} package to deal with pixelization, we can set a working resolution and translate sky direction $\hat{\bm n}$ to pixel $p$. The data model then becomes
\begin{equation}
    \bm S^{\tau,f} = \frac{4\pi}{N_{\rm pix}} \sum_p \bm \Gamma^{\tau,f}_{p} I_{p}^{}\,.
    \label{eq:data_model_I_pix}
\end{equation}
The components of Eq.~(\ref{eq:Imapsol}) can then be considered in the pixel domain, where $\hat{\bm n} \rightarrow p$.

\subsection{Amplitude-phase mapping}
Let us now take a step back and reconsider the data model in Eq.~(\ref{eq:data_f}), which we can write compactly as $\bm d^{\tau,f} =\bm s^{\tau,f} + \bm n^{\tau,f}$. Assuming zero-mean Gaussian noise in the strain, i.e., $\langle\bm n^{\tau,f}\rangle = 0$, Eq.~(\ref{eq:strain_model}) serves as our signal model.
As $\{h_+, h_\times\}$ are complex valued, there are four independent (real) fields on the sky to estimate:    $ \{h_+^\mathbb{R}, h_\times^\mathbb{R}, h_+^\mathbb{I}, h_\times^\mathbb{I}\}$.
We write down the likelihood in terms of the residuals in the strain as
\begin{equation}
    \mathcal{L} \propto \prod_{\tau, f, i, j} \frac{1}{|\bm C_n|^{1/2}} e^{-\frac{1}{2}\qty({\bm d} - {\bm s})^\dagger \, \bm ({\bm C_n})^{-1}\, \qty({\bm d} - {\bm s})} \,, 
\label{eq:Like}
\end{equation}
assuming each time segment $\tau$ and frequency $f$ in the dataset are statistically independent of one another, where we are again employing shorthand notation, using boldface font to represent a vector that now spans just the space of the detectors (and not baselines). $\bm C_n$ is the noise covariance $\bm C_n = \bm n^{} \otimes \bm n^\star$.
As the noise is modeled independently and is not part of the maximum likelihood estimation, it is possible to reduce the mapping problem to a closed-form $\chi^2$ solution minimizing
\begin{equation}
    \chi^2 = -\frac{1}{2} \sum_{\tau, f, i, j} (\bm d-\bm s)^\dagger({\bm C_n})^{-1}(\bm d- \bm s)\,.
\label{eq:chi2}
\end{equation}
We solve for the four $h$ fields separately; hence it is useful to decompose the signal model into real and imaginary components simply as
\begin{equation}
    \bm s^{\tau, f} = \int_{S^2} d\hat{\bm n} \, \sum_A \bm R^{\tau, f}_{A} (\hat{\bm n})\qty[\,h_A^{\mathbb{R}}(\hat{\bm n}) + i\,h_A^{\mathbb{I}}(\hat{\bm n})]\,,
\end{equation}
where the spectral dependence of the stochastic field has been explicitly factored out of $h_A(\hat{\bm n})$ and has been absorbed into the response term $\bm R$ so as to keep track of fewer dependencies. 
Minimizing $\chi^2$ with respect to each field yields the maximum likelihood solution
\begin{equation}
\colvec{2}{h_{A'}^\mathbb{R}}{h_{A'}^\mathbb{I}}_{\hat{\bm n}} =    \begin{pmatrix}
   M_{AA'}^\mathbb{R} & -M_{AA'}^\mathbb{I}\\
   M_{AA'}^\mathbb{I} & M_{AA'}^\mathbb{R}
   \end{pmatrix}^{-1}_{\hat{\bm n}, \hat{\bm n}'} \colvec{2}{z_{A'}^\mathbb{R}}{z_{A'}^\mathbb{I}}_{\hat{\bm n}'}\,,
\end{equation}
where 
\begin{equation}
     \colvec{2}{z_{A'}^\mathbb{R}}{z_{A'}^\mathbb{I}}_{\hat{\bm n}} = \sum_{\tau, f, i, j}   \colvec{2}{\bm z_{A'}^\mathbb{R}}{\bm z_{A'}^\mathbb{I}}^{\tau, f}_{\hat{\bm n}}\,,\qquad
   \begin{pmatrix}
   M_{AA'}^\mathbb{R} & -M_{AA'}^\mathbb{I}\\
   M_{AA'}^\mathbb{I} & M_{AA'}^\mathbb{R}
   \end{pmatrix}_{\hat{\bm n}, \hat{\bm n}'} = \sum_{\tau, f, i, j} 
   \begin{pmatrix}
   \bm M_{AA'}^\mathbb{R} & -\bm M_{AA'}^\mathbb{I}\\
   \bm M_{AA'}^\mathbb{I} & \bm M_{AA'}^\mathbb{R}
   \end{pmatrix}^{\tau, f}_{\hat{\bm n}, \hat{\bm n}'}\,.
    \label{eq:mapsol_sum}
\end{equation}
Each component is, explicitly,
\begin{equation}
    \begin{split}
        \bm z_A^\mathbb{R}(\hat{\bm n}) &= \bm R_{A}^{\mathbb{R}}(\hat{\bm n})\,{\bm C}_n^{-1}\, \bm d^\mathbb{R}+ \bm R_{A}^{\mathbb{I}}(\hat{\bm n})\,{\bm C}_n^{-1}\, \bm d^\mathbb{I}\,,\\
        \bm z_A^\mathbb{I}(\hat{\bm n}) &= \bm R_{A}^{\mathbb{R}}(\hat{\bm n})\,{\bm C}_n^{-1}\, \bm d^\mathbb{I}- \bm R_{A}^{\mathbb{I}}(\hat{\bm n})\,{\bm C}_n^{-1}\, \bm d^\mathbb{R}\,,\\
        \bm M_{AA'}^\mathbb{R}(\hat{\bm n}, \hat{\bm n}') &= \bm R_{A}^{\mathbb{R}}(\hat{\bm n})\,{\bm C}_n^{-1}\,\bm R_{A'}^{\mathbb{R}}(\hat{\bm n}')+\bm R_{A}^{\mathbb{I}}(\hat{\bm n})\,{\bm C}_n^{-1}\,\bm R_{A'}^{\mathbb{I}}(\hat{\bm n}')\,,\\ 
        \bm M_{AA'}^\mathbb{I}(\hat{\bm n}, \hat{\bm n}') &= \bm R_{A}^{\mathbb{R}}(\hat{\bm n})\,{\bm C}_n^{-1}\,\bm R_{A'}^{\mathbb{I}}(\hat{\bm n}')-\bm R_{A}^{\mathbb{I}}(\hat{\bm n})\,{\bm C}_n^{-1}\,\bm R_{A'}^{\mathbb{R}}(\hat{\bm n}')\,.
    \end{split}
    \label{eq:mapsol}
\end{equation}
For the sake of conciseness we do not include the frequency and time labels in each of the terms above; 
these mirror those in Eq.~\eqref{eq:Imapsol}. Note that the projection $z$ and Fisher matrix $M$ here are not the same as for intensity mapping---however, we have chosen to keep the same notation to draw the comparison between the two approaches.
In the uncorrelated noise case the (noise) covariance matrix becomes diagonal in detector space,
\begin{equation}
    ({\bm C}_n)_{ij} = \delta_{ij} P_j
\end{equation}
so calculations in Eq.~(\ref{eq:mapsol}) simplify considerably.
Furthermore, as in the intensity mapping approach described above, the integration over frequencies in Eq.~\eqref{eq:mapsol_sum} requires an assumption for the spectral shape of the signal. However, in the amplitude-phase case one must take extra care: when broadband integrating here one must assume a certain phase-coherence across modes, or else the phase introduced in Eq.~(\ref{eq:phasestrain}) will (correctly) average to zero and yield null sky maps. 

Finally, as in the previous section, we can discretize the sky and transform sky direction $\hat{\bm n}$ to pixel $p$, to apply this method to data. The signal model becomes
\begin{equation}
    \bm s^{\tau,f} = \frac{4\pi}{N_{\rm pix}} \sum_p\sum_A \bm R^{\tau,f}_{A,\,p} h_{A,\,p}^{}\,,
    \label{eq:data_model_pix}
\end{equation}
and similarly the terms in Eq.~(\ref{eq:mapsol}) may be translated into the pixel domain.

\section{Instructive Considerations}\label{sec:toys}
Before discussing the application of the map-making methods laid out above, let us start with some useful considerations about the nature of the measurement of broadband, stochastic GWs from all sky directions with a set of interferometers.
In particular, let us focus on the notion that a detector naturally low-pass filters the signal, both in terms of temporal resolution and angular resolution; the subsequent digitization/sampling of the data needs to respect the maximum temporal and spatial frequencies present in the signal if aliasing of power is to be avoided.

It is useful to first break down the measurement into two components: the time or frequency-domain measurement, and the sky-dependent response. These two steps are inextricable in a real GW detector. However, to get an idea of the accessibility to each information component we can consider two simple scenarios separately: first, we can focus on the estimation of the amplitude and phase of a complex field in the frequency domain through time-domain measurements; then we can analyze the effect of a window-averaged response on the measurements of a complex field on the semicircle, in analogy with what happens with GW detectors observing the full sky.

When analyzing GW detector timestreams that contain a measurement of a stationary stochastic field, say $h(t)$, we typically start by taking the Fourier transform (FT) of the data, as the frequency domain representation allows us to access information more efficiently. A way to see this is to note that the quadratic expectation value in the time and frequency domains are related by
\begin{equation}
    \langle h(t) h(t') \rangle = C(t-t') \quad \xleftrightarrow{\rm FT}{} \quad\langle \tilde{h}(f) \tilde{h}^\star(f') \rangle = P(f) \delta(f-f')\,.
\end{equation}
Thus, the Fourier transform of a stationary timeseries maps our measurement to the diagonal space for the autocorrelation of the field, where $P(f)$ is the (two-sided) power spectrum of the field. In other words, while in the time domain the correlation between the stationary field at different times depends (only) on the time difference $t-t'$, in the {\it conjugate space} (frequency space) the correlation between different frequency components is a delta function, and thus the frequency dependence is totally ``compactified'' in the power spectrum. We can either write an estimator for a real field, for example, $P(f)$, if we are interested only in the intensity of the signal, or we can aim to measure the complex field $\tilde{h}(f)$, to include both amplitude and phase information. In either case, we model the signal in the frequency domain and estimate it through time-domain measurements. Thus, in practice the measurement of the field is performed in the conjugate space of the model, over an observation time $T$, sampled at the finite time resolution $\Delta t_d$. According to the Nyquist–Shannon sampling theorem, given the Nyquist frequency of the measurement $f_{\rm Nyq} \equiv 1/(2\Delta t_d)$, the field $\tilde{h}(f)$ may be completely determined as long as it is made up of modes with frequency content $1/\Delta t<f_{\rm Nyq}$, where $\Delta t$ is the time-coherence scale of the field.
Hence, {\it if} the GWs in the timestream are at frequencies below $f_{\rm Nyq}$, it is theoretically possible to fully reconstruct the Fourier coefficients $h_A(f)$, in the case of an isotropic background. 
If the target signal is stochastic in nature, and the phase information is simply uninteresting, then the choice of method boils down to what is most computationally effective when dealing with noisy timestreams and multiple detectors.

To understand the effects of the integrated beam of GW detectors, let us start by considering a complex and statistically isotropic field $\tilde{h}(\hat{\bm n})$ on the sphere,
\begin{equation}
\tilde{h}(\hat{\bm n}) = {\cal A}(\hat{\bm n}) e^{i\varphi(\hat{\bm n})}\,,
\end{equation}
where the phase $\varphi$ has a certain functional dependence on direction. This is effectively a frequency-independent, unpolarized version of the GW strain in the Fourier domain given by Eq.~(\ref{eq:phasestrain}). Note that, as in the relation $t\xleftrightarrow{\text{FT}}{}f$, there is a useful conjugate space to direction $\hat{\bm n}$ space,
\begin{equation}
    \langle \tilde{h}(\hat{\bm n}) \tilde{h}^\star(\hat{\bm n}') \rangle = C(\hat{\bm n}\cdot\hat{\bm n}') \quad \xleftrightarrow{\rm SHD}{} \quad\langle a_{\ell' m'} a_{\ell' m'} \rangle = C_\ell \delta_{\ell\ell'}\delta_{mm'}\,,
    \label{eq:shd_decomp}
\end{equation}
where $a_{\ell m}$ are the spherical harmonic coefficients of $\tilde{h}(\hat{\bm n})$, and $C_\ell$ is the angular power spectrum of $\tilde{h}$. The $C_\ell$s are related to $C(\hat{\bm n}\cdot\hat{\bm n}')$ via
\begin{equation}
    C(\hat{\bm n}\cdot\hat{\bm n}') = \sum_{\ell=0}^\infty \frac{2\ell+1}{4\pi} C_\ell P_\ell(\hat{\bm n}\cdot\hat{\bm n}')\,,
\end{equation}
where $P_\ell(x)$ are the Legendre polynomials.  
The spherical harmonic decomposition (SHD) introduced in Eq.~\eqref{eq:shd_decomp} allows us to transform the field from directional space $\hat{\bm n}$ to degree $\ell$ and order $m$ space, and, again, the correlation is diagonal in the latter.
The measurement of $\tilde{h}$, however, does not occur in $\hat{\bm n}$ space directly, but in the time domain; the relation between these two spaces depends on the specific detector used to make the measurement. This is the crucial difference between measurements of a frequency-dependent GW observable and a direction-dependent one. 

Let us now discuss an instructive example. Let us consider a case where we can fix the observation to a single value of $\phi$ on the sphere, such that we reduce the mapping problem to the estimation of a one-dimensional field on the semicircle. $\tilde{h}$ is then parametrized solely by $0\leq\theta\leq\pi$ on the semicircle, such that effectively we are observing $\tilde{h}(\theta, \phi={\rm const.})\equiv \tilde{h}(\theta)={\cal A}(\theta) e^{i\varphi(\theta)}$.
We assume that our detector makes a time-dependent measurement $r(t)$ of the field on the semicircle, 
\begin{equation}
    r(t) = \int_{0}^\pi d\theta R(t, \theta)\, \tilde{h} (\theta)\,,
    \label{eq:Phi_d}
\end{equation}
filtered by the detector response $R(t,\theta)$, which scans the semicircle as time goes by. Note here the similarity with the model for the GW strain in Eq.~(\ref{eq:strain_model}), relevant to the measurement of anisotropic GWBs.

The intrinsic coherence scale of the field, $\Delta\theta$, sets the input resolution of our example, inducing ${\pi}/{\Delta\theta}$ independent samples on the semicircle. We take the phase $\varphi(\theta)$ to be random at the intrinsic scale, such that the field has statistically independent phases at each sample. The amplitude $\cal A$ is assumed to vary much more slowly than the phase, such that the scale over which the amplitude varies appreciably is much larger than $\Delta\theta$.
We model the instantaneous response of the detector at time $t_0$ as a top-hat function with width $\Delta\theta_d$,
\begin{equation}
    R(t_0, \theta) = \left\{ \begin{array}{ll}
      1\,, & -\frac{{\Delta}\theta_d}{2}<\theta (t_0)<\frac{{\Delta}\theta_d}{2} \\
      0\,, & \text{otherwise} \\
\end{array} 
\right. \,,
\end{equation}
and the time dependence is imposed by the scanning strategy, i.e., the angular function $\theta(t)$. In this simple example we take $\theta (t) = \pi\, t/\tau$ where $\tau$ is the time period over the semicircle, such that consecutive measurements correspond to consecutive angular samples on the semicircle, and continuity between measurements at 0 and $\pi$ is ensured.
The measurement of $\tilde{h}$ is then mediated by the detector-induced window width ${\Delta}\theta_d$, and the comparison between $\Delta\theta$ and ${\Delta}\theta_d$ will determine to what degree the field is resolvable. ${\Delta}\theta_d$ is the {\it angular resolution} of the measurement, and as long as ${\Delta}\theta_d$ is sufficiently less than $\Delta\theta$, the field is overresolved and its amplitude and phase information may be measured exactly. However if ${\Delta}\theta_d > \Delta\theta$ there is loss of information as the window averages over the phases, which are randomly distributed along the semicircle. In the extreme case where ${\Delta}\theta_d \gg \Delta\theta$, the measurement becomes compatible with 0, and the amplitude of the field is entirely lost. 
Translating the measurement into the conjugate space, one may interpret this limit as an \emph{angular} Nyquist frequency, $\ell_{\rm Nyq} \equiv \pi/(2{\Delta}\theta_d)$, imposed by the detector. In this sense, the measurement only works when $\ell_{\rm Nyq}>\ell_{\rm max}$, where $\ell_{\rm max} \sim \pi/\Delta\theta$ is the maximum angular frequency of the signal. Note the analogy then between ${\Delta}\theta_d$, and the sampling rate $\Delta t_d$ described above.
The useful estimator for this example is one that marginalises over the phases and aims for the field intensity  $C(\theta)$ directly, which is
\begin{equation}
    \langle \tilde{h}^{}(\theta) \tilde{h}(\theta')\rangle = C\qty(\cos(\theta)) \delta(\theta-\theta')\,,
    \label{e:Idef}
\end{equation}
obtained from Eq.~\eqref{eq:shd_decomp} assuming total phase incoherence of $\tilde{h}(\theta)$.
Equal-time measurements $r_1(t)$, $r_2(t)$ made by two identical detectors $1$, $2$ then satisfy
\begin{equation}
    \langle r^{}_1(t) \, r^\star_2(t)\rangle = \int_{0}^\pi d\theta R^{}_1(t, \theta) R_2^{\star}(t, \theta) C(\cos(\theta))\,, 
    \label{eq:toyImapping}
\end{equation}
where the angle brackets introduce an expectation value similar to Eq.~(\ref{eq:stokey}); i.e., they imply ensemble averaging over different data samples, and the Dirac delta in Eq.~\eqref{e:Idef} has already been applied.
Mapping then amounts to inverting Eq.~(\ref{eq:toyImapping}) to estimate $C(\theta)$. Hence to preserve the amplitude information it is necessary to cross-correlate (i.e., square) signals before averaging over the response window. Note that the spatial averaging introduced in Eq.~(\ref{eq:Phi_d}) is analogous to what happens in a GW detector: the directional information is not directly accessible in the measurement, but rather needs to be reconstructed based on observation features, such as the time dependence of $R$. 

In the simple case above the only relevant coherence scale of the field is that of the phase; however, in general it is also necessary to worry about the coherence scale of the intensity, $\Delta \theta_C$. Note this is directly related to the coherence scale of the amplitude, as in this case $C ={\cal A}^2$. When $\Delta \theta_C<{\Delta}\theta_d$, this will have an impact on the observation of $C$ (or, equivalently, of ${\cal A}$, taking the square root of the intensity) similar to what occurs for $\varphi$. However, note that as the intensity is positive definite, this field will never average to zero but would rather approach zero from above in the limit of $\Delta \theta_C\ll{\Delta}\theta_d$.

\section{\sl Mock-data Mapping Tests}\label{sec:tests}
Simplified mock-data mapping tests are presented here to illustrate the points made in the previous sections. We consider the two LIGO interferometers that have a well-known response function, illustrated, for example, in~\cite{Romano2017}. In these examples, the noise is taken to be white, Gaussian, stationary, and much weaker than the signal. This is to highlight the potential and shortcomings of the mapping algorithms based on the characteristics of the signal component only. As the noise is assumed to be known, it is not part of the estimation. In practice, all GW mapping attempts to date using ground-based detectors have relied on independent estimators for the noise, which is a valid approach when the signal is entirely subdominant and hence does not bias the noise estimation, at least on a segment-by-segment basis. This methodology needs to be revisited in the presence of competing signal and noise components---an example of this may be seen in~\cite{PhysRevD.102.043502} for the Laser Interferometer Space Antenna.

The reconstructions presented here are obtained following the data-handling recipe of~\cite{Renzini:2018vkx} which imitates the steps performed on real LIGO data~\cite{LIGOS4,LIGOS5}. In both intensity and amplitude-phase mapping, the mock data are generated in segments that represent Fourier transformed one-minute segments of a detector array time-stream. The one-minute time scale is chosen such that the response of the detectors may be considered constant on the sky throughout the segment. This not only sets a natural lower bound on the frequency range probed, but also imposes a pixelization scheme: in order for the reconstruction to be complete, the sky response should vary smoothly from segment to segment. 

The most challenging part of this procedure is the safe inversion of the Fisher matrix. This arises since the spatial sampling of the sky is suboptimal, as explored, for example, in~\cite{Renzini:2018vkx}. However, as long as the noise is subdominant it is possible to pseudo-invert $M$ discarding a minimal number of modes and recovering the sky signal perfectly, as shown below. The situation capsizes in the presence of noise: in that case, it is necessary to choose a cutoff on the singular values of the eigenmodes of $M$, to differentiate between the signal and noise modes. This is a highly nontrivial problem, so we do not discuss it here---see the discussion in~\cite{Abbott:2021jel} for example.    

In all mapping examples presented here, the signal is modelled as a diffuse point source on the sky, with varying phase as a function of direction. This is obtained in practice by associating a random phase to each pixel of the amplitude sky map, which is then scanned to prepare the data segments. This is explicitly performed via Eq.~(\ref{eq:data_model_pix}) in the case of amplitude-phase mapping, and Eq.~(\ref{eq:crossdata_model_pix}) in the case of intensity mapping. The pixelization schemes used for data generation and map reconstruction need not match---in fact, one may not in general expect that the natural coherence scale of the field be comparable to that of the detector. The size of the unit pixel at injection,~$4\pi / N_{\rm pix}^{\rm in}$, is analogous to $\Delta\theta$ in the example made in Sec.~\ref{sec:toys}, while the size of the unit pixel at reconstruction,~$4\pi / N_{\rm pix}^{\rm out}$, is analogous to ${\Delta}\theta_d$. For simplicity, we do not add a frequency-dependent phase term, although in reality it is present and may not be neglected in a real GW measurement; we discuss this further below.


\begin{figure}
    \centering
    \includegraphics[width = 0.9\textwidth]{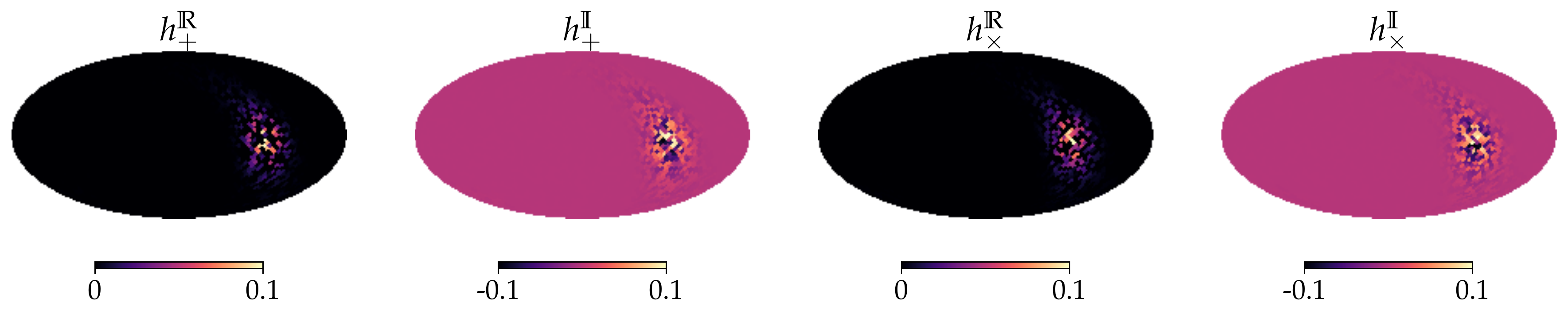}
    \includegraphics[width = 0.4\textwidth]{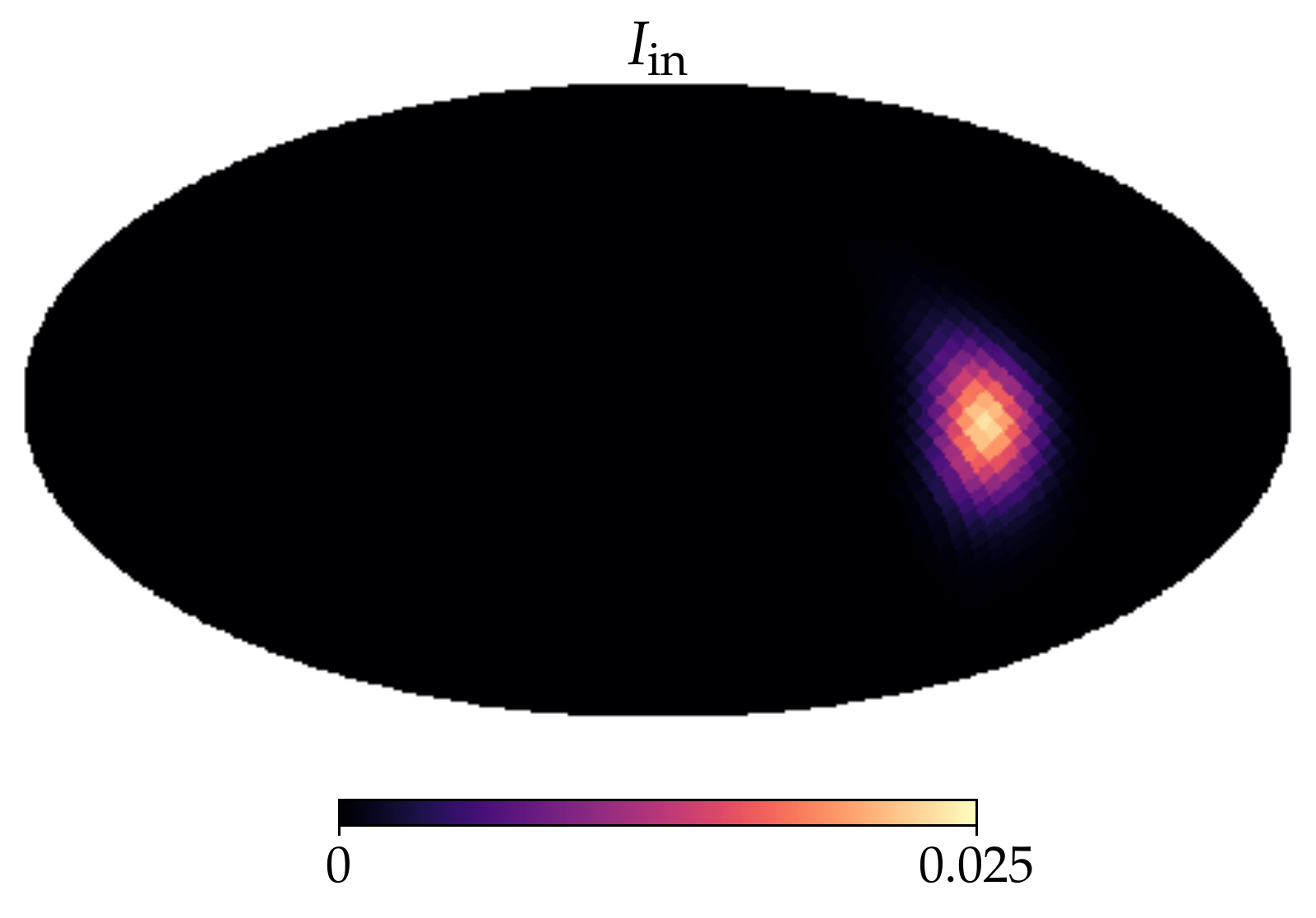}
    \caption{Set of input sky maps used in the mock-data tests described in the text. The top row shows injected maps for the four components of the $h$ field, used for data injection and reconstruction with the amplitude-phase algorithm. The pixel phase is chosen randomly pixel to pixel. The bottom map corresponds to the total injected intensity $I_{\rm in}$ resulting from the combination of the maps in the top row. All these maps are injected at $N^{\rm in}_{\rm side}=16$.}
    \label{fig:GB_point_source_in}
\end{figure}

\begin{figure}
    \centering
    \includegraphics[width = 0.9\textwidth]{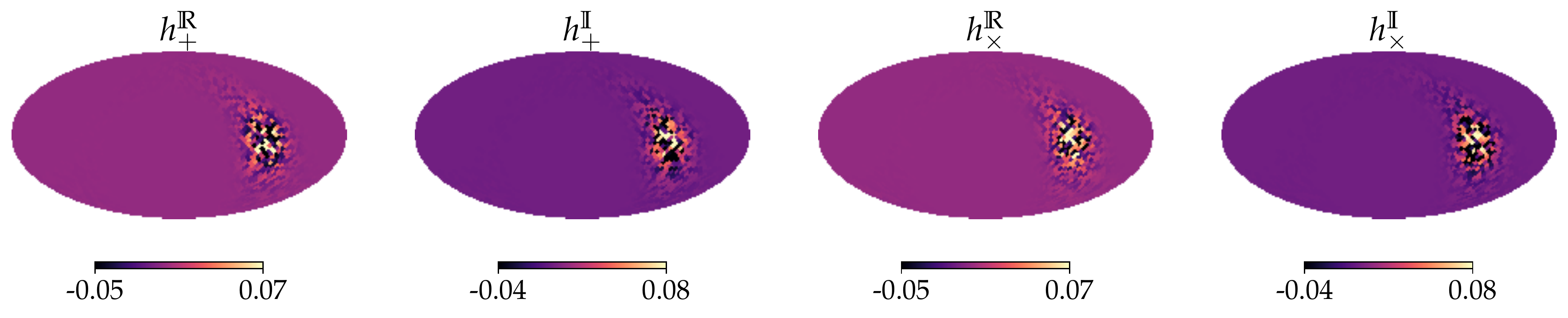}
    \includegraphics[width = 0.9\textwidth]{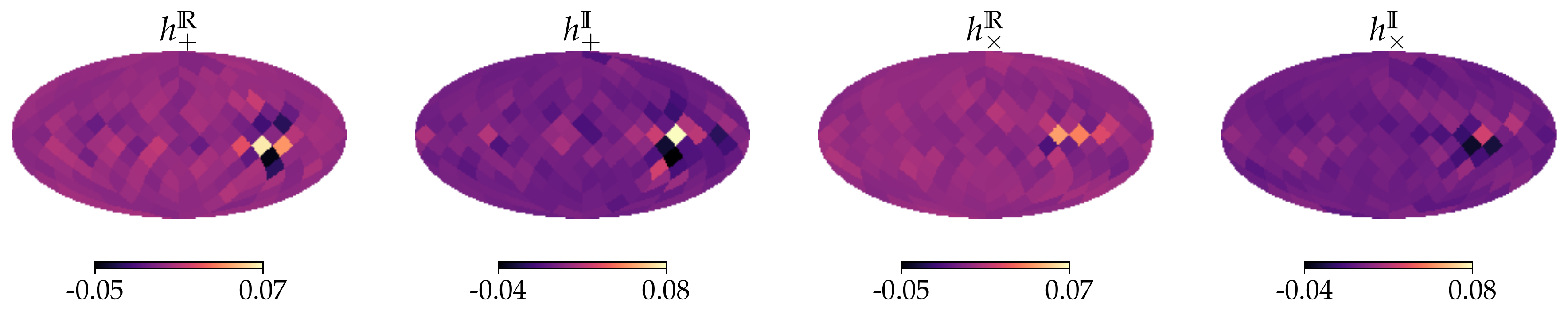}
    \includegraphics[width = 0.9\textwidth]{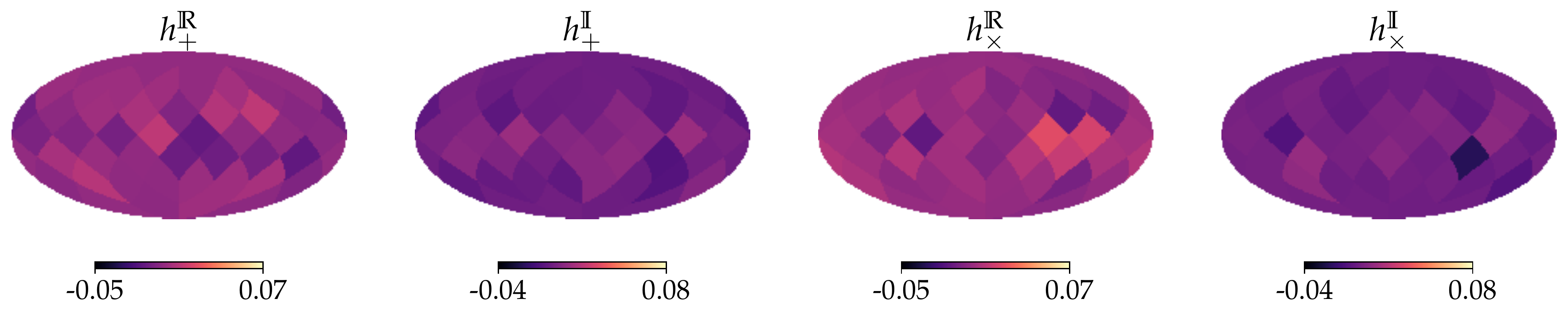}
    \includegraphics[width = 0.8\textwidth]{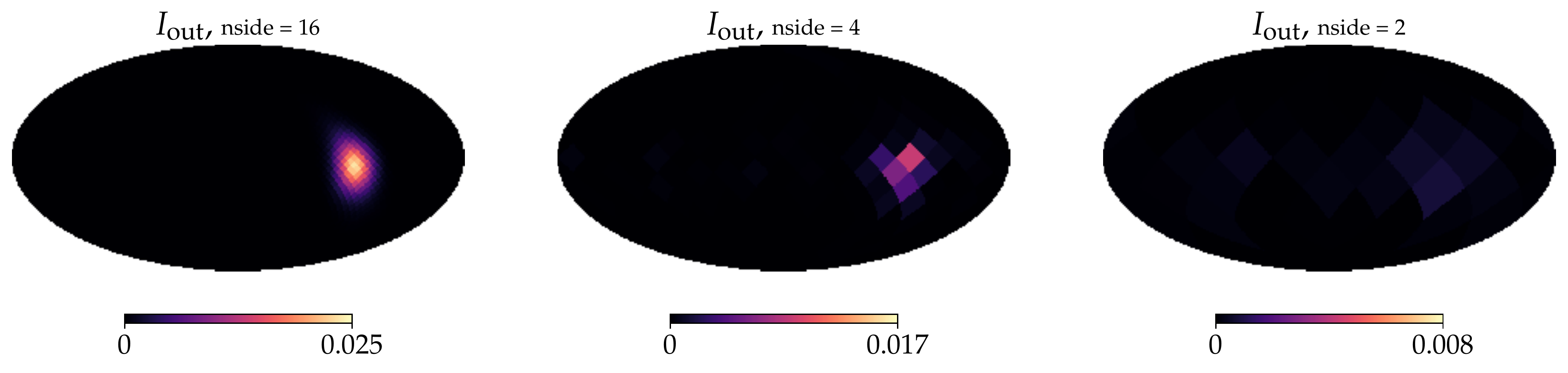}
    \caption{
    Set of sky map reconstruction examples with the amplitude-phase algorithm, using maps shown in Fig.~\ref{fig:GB_point_source_in} as input.
    The top row shows reconstructed maps with $N_{\rm side}^{\rm out} = 16$, which corresponds to the case when the intrinsic coherence scale of the field is equal to that of the detector.
    The second and third rows show reconstructions with $N_{\rm side}^{\rm out} = 4$ and $N_{\rm side}^{\rm out} = 2$, respectively. The bottom row shows the corresponding recovered intensity maps for the three different $N_{\rm side}$ cases. In the case when $N_{\rm side}^{\rm out} = N_{\rm side}^{\rm in}$, the recovered intensity matches exactly the one shown in Fig.~\ref{fig:GB_point_source_in}; in the other cases, the intensity is degraded.}
    \label{fig:GB_point_source_out}
\end{figure}

Fig.~\ref{fig:GB_point_source_out} shows results obtained with the amplitude-phase mapping algorithm, using the input maps from Fig.~\ref{fig:GB_point_source_in}. The top row of Fig.~\ref{fig:GB_point_source_out} shows the case when the input and output resolutions match exactly, $N^{\rm in}_{\rm side} \equiv N^{\rm out}_{\rm side} = 16$; hence the phases of the $h_A$ fields in each pixel are perfectly recovered. However, the second and third rows of Fig.~\ref{fig:GB_point_source_out} show two cases where the output resolution is lowered from the input $N^{\rm in}_{\rm side} = 16$ to $N^{\rm out}_{\rm side} = 4$ and $N^{\rm out}_{\rm side} = 2$ respectively; here, the phase components are averaged out within neighboring pixels, hence the phase information is lost. This leads to an overall loss of intensity information as well, by the same reasoning described for the example in Sec.~\ref{sec:toys}. Note that style choices have been made to underline the effect: the values chosen for the color bars in the intensity $I$ plots in Fig.~\ref{fig:GB_point_source_out} match the correct values one would obtain when coarse-graining the $I_{\rm in}$ intensity map to the relevant $N_{\rm side} ^{\rm out}$. The coarse-graining exercise is performed with the {\tt ud\_grade} function of the {\tt healpy} package~\cite{Gorski2004}.

\begin{figure}
    \centering
    \includegraphics[width = 0.9\textwidth]{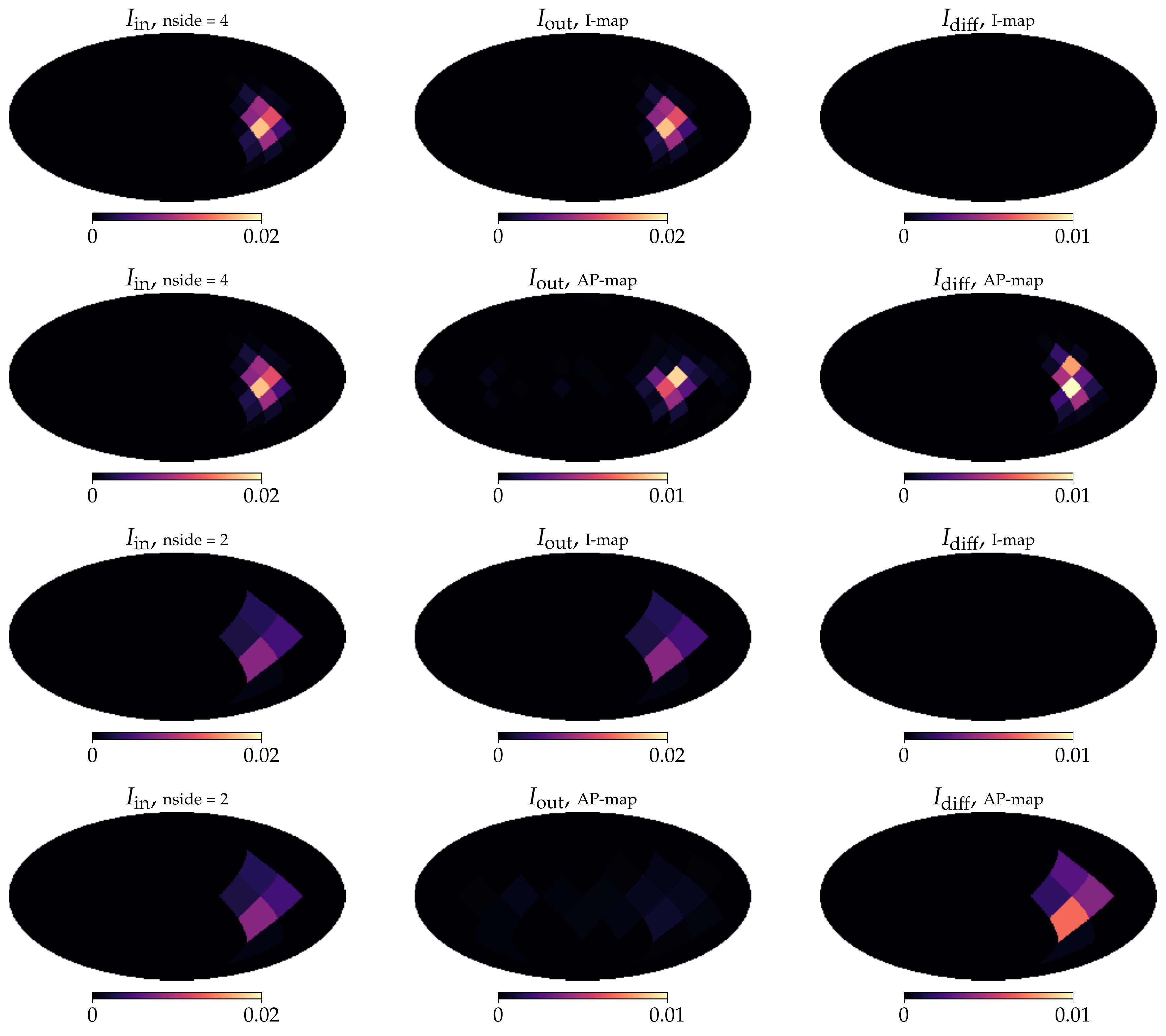}
    \caption{A comparison between the recoveries of the same injected map $I_{\rm in}$ with the intensity $I$ (first and third rows) and amplitude-phase $AP$ (second and fourth rows) mapping algorithms, with the two different values of $N_{\rm side}^{\rm out}$ as shown in Fig.~\ref{fig:GB_point_source_out}. On the right, the difference maps $I_{\rm in} - I_{\rm out}$ are shown to quantitatively comparative the performance of the algorithms. Note that the input map has been {\tt ud\_grade}d to allow for this comparison; the actual input map is at $N_{\rm side} = 16$ as shown in Fig.~\ref{fig:GB_point_source_in}.}
    \label{fig:GB_diff}
\end{figure}

This effect does not occur in the case of intensity mapping, since the phase information is not required to reconstruct the intensity on the sky. We have repeated the exercise above using the same input map as in Fig.~\ref{fig:GB_point_source_in} employing the intensity mapping algorithm, varying the values of $N_{\rm side} ^{\rm out}$, and find that the $I_{\rm out}$ maps agree with the {\tt ud\_grade}d input maps within less than $1\%$.  To quantitatively compare intensity and amplitude-phase mapping, we show in Fig.~\ref{fig:GB_diff} the difference between the recovered maps with intensity mapping, $I^{}_{\rm out, I-map}$, and amplitude-phase mapping, $I^{}_{\rm out, AP-map}$, in the cases where $N_{\rm side}^{\rm out}<N_{\rm side} ^{\rm in}$. In both cases, it is clear that most of the overall power is lost when performing amplitude-phase mapping, and in particular in the case where $N_{\rm side}^{\rm out}=2$, the injected point source is hardly discernible from the fluctuations around 0 in the pixels.

\begin{figure}
    \centering
    \includegraphics[width = 0.8\textwidth]{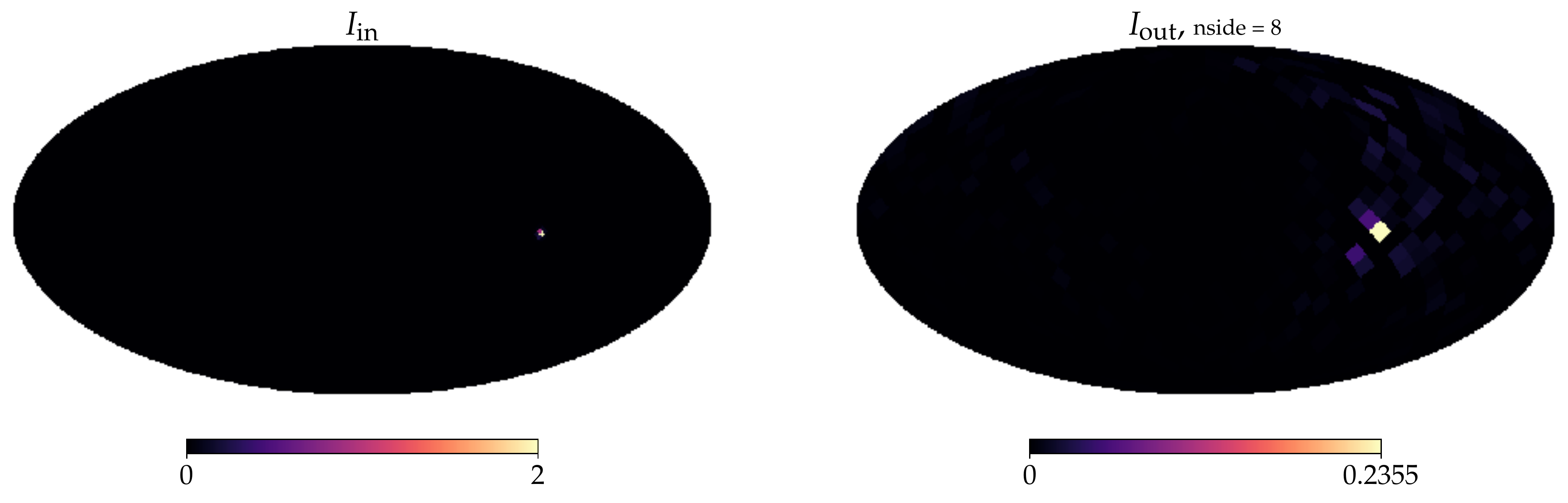}
    \caption{Sky map injection (left panel) and reconstruction (right panel) with the intensity mapping algorithm, in a case where the intrinsic coherence scale of the signal intensity is considerably smaller than that of the detector.}
    \label{fig:GB_point_source_I}
\end{figure}

Hence, we confirm that in the case of intensity mapping, the overall GW intensity is conserved, and the structure on the sky as well, as long as the coherence scale of the intensity is larger than that of the detector. In Fig.~\ref{fig:GB_point_source_I}, we show an example where this is not the case: here, the injected signal is a very high resolution input map, $N^{\rm in}_{\rm side} = 32$, which is null everywhere apart from a tiny patch, mimicking a pointlike source, and is recovered at $N^{\rm out}_{\rm side} = 8$. Here indeed the signal from the (almost unique) incoming direction is averaged within the neighboring pixels, and hence the point value in that specific direction is much lower than the injection (by a factor of $\sim 10$). Note, however, that the recovery is perfectly in line with a coarse-graining of the input map: the monopole is conserved in the operation, and the correct value is recovered in the single ``hot'' pixel. Hence, in this case there is as little loss of information as possible. But this may present an issue when the signal is competing with a high level of noise.

Adding frequency-dependent phase terms to the injected maps has no effect on the intensity mapping results, as the data are first cross-correlated in the frequency domain and the phase term cancels out. However, in the amplitude-phase case this adds another level of complexity, as we need to keep track of the individual mode phases as well as those in each pixel. In fact, each frequency bin is independent, given the assumption of time stationarity, hence band-integrating over frequency bins would then have a similar effect as attempting map-making with a detector resolution that is larger than the intrinsic coherence scale of the field: the signal is degraded as the recovered amplitude averages to zero with the number of bins. 

\section{\sl Discussion}\label{sec:discussion}

We have taken a pedagogical approach in discussing the advantages and shortcomings of two different maximum-likelihood map-making techniques---intensity mapping and amplitude-phase mapping. The intent was to demystify these techniques and clarify the potential of detection and mapping of GWBs in general. 
Intensity mapping targets the GWB intensity as a function of direction. It constrains the average GW power on the sky and any anisotropies about this average. Versions of the method presented here have been applied to LIGO--Virgo data and have been proposed for LISA~\cite{Cornish:2001hg,Cornish:2002bh,PhysRevD.92.042003,banagiri2021mapping}. This method is best suited to astrophysical stochastic backgrounds, which carry no phase information and hence solving for it would unnecessarily complicate the mapping procedure. On the other hand, amplitude-phase mapping allows one to do just that; however, we have shown that it is only possible to resolve the phase component if the intrinsic angular coherence scale of the signal is comparable to or larger than the angular resolution of the detector. When this is not the case, attempting to estimate the phase on the sky leads to loss of information, including the amplitude of the signal. Hence, this method is to be avoided in the case of stochastic backgrounds, where the intrinsic coherence scale is usually zero. 

A possible exception to this rule may be the case of a stochastic background dominated by very few sources, such as that considered in~\cite{cornish2014mapping}. This is a possibility for the stochastic background probed by pulsar timing arrays (PTAs), as the signal should be dominated by $\sim10^2-10^3$ sources on the sky, such that a single source should dominate a pixel of $\sim 40$ deg$^2$, which may be taken as the intrinsic coherence scale of the background. Even in this case, we are not quite there yet: achieving such an angular resolution with signal to noise ratio SNR$\sim$3 would require a pulsar array with almost $10^4$ pulsars~\cite{PhysRevD.90.082001}, while current PTAs are monitoring $\sim10^2$ pulsars, setting the present resolution to $\sim400$ deg$^2$. Hence, this remains a target for future observatories, such as the Square Kilometre Array.

We have provided a useful investigation into the relationship between the time--frequency and the pixel--angular frequency domains. Time samples $t$ are conjugate to frequencies $f$, while pixels $p$ which correspond to sky locations $\hat n$ are conjugate to the angular ``frequency" scale $l$. However, our sky observations are effectively carried out in the time domain. A detector naturally low-pass filters the data, both in terms of temporal resolution and angular resolution; however, while the temporal resolution is  well-known and controlled by the experimenter, the angular resolution is determined by the geometry and motion of the detector array. Furthermore, the digitization and sampling of the data needs to respect the maximum temporal and spatial frequencies, if aliasing of power is to be avoided.

The first detection of a GWB is drawing near, whether it be with PTAs, possibly by the NANOGrav Collaboration~\cite{Arzoumanian:2020vkk}, or with the LIGO--Virgo--Kagra ground-based interferometer network. Currently, searches are focusing on the isotropic background, as astrophysical backgrounds are not expected to be highly anisotropic~\cite{Cusin:2017fwz, Jenkins:2019nks}. However, understanding how to handle any anisotropy to confirm or rule out this hypothesis remains an essential task of our collaborations. Ideally, the isotropic GWB could be estimated as the monopole of GWB maps~\citep[as in Ref.][for example]{Renzini2019b}, and not under the strict assumption of zero anisotropy. 

\section{Acknowledgments}
A.I.R. acknowledges the support of the National Science Foundation and the LIGO Laboratory.
J.D.R. acknowledges support from NASA grant 80NSSC19K0318, NSF Physics Frontiers Center Awards No. PFC-1430284 and No. PFC-2020265 and start-up funds from Texas Tech University. C.R.C. acknowledges support by Science and Technology Facilities Council consolidated Grant No. ST/P000762/1. N.J.C. appreciates the support of NASA LISA foundation Science Grant No. 80NSSC19K0320, NSF Award No. PHY1912053, and NSF Physics Frontiers Center Awards No. PFC-1430284 and No. PFC-2020265.

\bibliographystyle{apsrev}
\bibliography{refs.bib}

\end{document}